\title{A Digital Twin Framework for Reinforcement Learning with Real-Time Self-Improvement via Human Assistive Teleoperation}
\documentclass[lettersize,journal]{IEEEtran}
\usepackage{amsmath,amsfonts}
\usepackage{algorithmic}
\usepackage{algorithm}
\usepackage{array}
\usepackage[caption=false,font=normalsize,labelfont=sf,textfont=sf]{subfig}
\usepackage{textcomp}
\usepackage{stfloats}
\usepackage{url}
\usepackage{verbatim}
\usepackage{graphicx}
\usepackage{cite}
\usepackage{array}
\begin{document}

\title{A Digital Twin Framework for Reinforcement Learning with Real-Time Self-Improvement via Human Assistive Teleoperation}

\author{Kabirat Olayemi, Mien Van*, Luke Maguire, Sean McLoone~\IEEEmembership{Senior Member,~IEEE}
        % <-this % stops a space
\thanks{This work was supported by The Department for Education, Northern Ireland.}% <-this % stops a space
\thanks{The authors are with The Centre for Intelligent Autonomous Manufacturing Systems, School of Electronics, Electrical Engineering and Computer Science, Queens University, Belfast, UK. {\tt\small \{kolayemi01, m.van(*Corresponding author),lmaguire32, s.mcloone\}@qub.ac.uk.}}} 

\maketitle

\begin{abstract}
Reinforcement Learning (RL) or Deep Reinforcement Learning (DRL) is a powerful approach to solving Markov Decision Processes (MDPs) when the model of the environment is not known a priori. However, RL models are still faced with challenges such as handling covariate shifts and ensuring the quality of human demonstration. To address these challenges and further advance DRL models, our work develops a human-in-the-loop DRL framework via digital twin that leverages human intelligence after deployment to retrain the DRL model in real time. First, we develop a pre-trained model fully based on learning through trial and error in the simulated environment allowing scalability and automation while eliminating variability and biases that can come from subjective human guidance. Second, instead of deploying the trained model directly on the UGV, we create a digital twin which controls the physical UGV from the virtual environment. Third, to allow continuous learning without catastrophic forgetting, we introduce the ability of the model to self-improve with the help of small human guidance at the start of the retraining. We test the performance of our proposed model in both simulation and real-world environments with both static and dynamic obstacles. The results indicate that our proposed approach not only outperforms the baseline models in terms of reward accumulation but also demonstrates superior training efficiency.
\end{abstract}

\begin{IEEEkeywords}
Unmanned ground vehicles, digital twin, reinforcement learning, human-in-the-loop, TD3, navigation
\end{IEEEkeywords}

\section{Introduction}
\IEEEPARstart{U}{nmanned} ground vehicles (UGVs) play a crucial role in various sectors including military surveillance, agricultural mechanisation, industrial automation, urban logistics, emergency management, and infrastructure management \cite{hu2023use, lv2021research}. 

Typically, the navigation system for UGVs is composed of several interdependent modules, each responsible for a specific aspect of navigation \cite{Nguyen}. These modules include perception, localization, mapping, path planning, motion control, and system integration \cite{rone2013mapping,fethi2018simultaneous,wu2020robust}. The perception module is the cornerstone of a UGV’s navigation system, responsible for interpreting the environment using a suite of sensors such as LiDAR, cameras, radar, and IMUs. The localisation module builds on the information provided by the perception module to determine the UGV's precise position and orientation within the environment using technologies such as SLAM, visual odometry, and odometry. The mapping module uses the localization data to create and update a map of the environment such as grid maps, topological maps, and 3D maps using techniques such as occupancy grids and probabilistic mapping, which is essential for path planning and obstacle avoidance. 

With an up-to-date map provided by the mapping module, the path planning module computes a collision-free route from the UGV’s current location to its target destination. This module employs global path planning algorithms like A*, Dijkstra, and D* \cite{yang2016survey}, as well as local planning methods such as the Dynamic Window Approach (DWA) and Timed Elastic Band (TEB) \cite{shen2023targeted}. The motion control module then translates the planned path into executable commands for the UGV’s actuators, ensuring smooth and precise movement. Using control algorithms such as PID controllers and Model Predictive Control (MPC), this module adjusts the UGV’s velocity and steering to follow the planned trajectory. Finally, system integration ensures that all the navigation modules work together seamlessly using middleware platforms like ROS (Robot Operating System) to facilitate communication and data handling between modules, while simulation environments such as Gazebo provide a testing ground to validate the integrated system before real-world deployment. 

However, this approach is faced with many challenges such as (1) managing drift and error accumulation over time in the localization module \cite{cadena2016past}, (2) handling the dynamic nature of real-world environments requires the mapping module to continuously update its representations while also efficiently handling the storage and processing of large maps \cite{macario2022comprehensive}, and (3) maintaining dynamic stability, especially on varied terrains, and ensuring precise actuator control linking the abstract plans to physical actions \cite{aggarwal2020path}. 

The above-mentioned challenges have prompted researchers to dive into the use of deep reinforcement learning (DRL) or reinforcement learning (RL) \cite{sivashangaran2021intelligent} for the data-driven navigation method. These methods provide a framework for UGVs to learn optimal navigation strategies through interaction with their environments. Thus, the DRL or RL algorithms can enhance the accuracy of localization by learning to fuse multiple sources of data (e.g., GPS, IMU, visual odometry) more effectively for better position and orientation estimates. It does not need to pre-create the environment map beforehand but rather employs the changing values in the sensor information, and it helps UGV's to learn to approximate optimal paths quickly while adapting to real-time changes in the environment. Despite this significant improvement, DRL/RL navigation models face several challenges.

Firstly, developing DRL/RL models for UGVs involves extensive testing and validation to ensure reliability and safety in diverse environments which is impractical in real-world scenarios. Secondly, simulation-based training has become a key approach to solving the first problem, allowing cost-effective, safe training and accelerated testing while mitigating risks associated with real-world trials before deployment. Nevertheless, the challenge of adapting these algorithms for real-world applications—known as sim-to-real transfer—continues to be a substantial hurdle due to the discrepancies between simulated and real environments \cite{9468918}. Thirdly, ensuring that RL policies perform reliably under varying and unforeseen conditions is difficult. RL policies may fail in scenarios that were not adequately represented during training \cite{henderson2018deep}. Lastly, implementing continuous learning in DRL models without catastrophic forgetting (where new learning disrupts previously acquired knowledge) is also a significant challenge \cite{kirkpatrick2017overcoming}.

To address the challenges mentioned above and further advance DRL models, our work develops a digital twin framework for RL real-time self-improvement via human assistive teleoperation that leverages human intelligence after deployment to retrain the DRL model in real-time. First, we develop a pre-trained model fully based on learning through trial and error in the simulated environment allowing scalability and automation while eliminating variability and biases that can come from subjective human guidance. Second, rather than deploying the trained model directly on the UGV, we created a digital twin that controls the physical UGV from the virtual environment based on the information received from the physical UGV. This helps to address the problem with sim-to-real transfer. Third, to allow continuous learning without catastrophic forgetting, we introduce the ability of the model to retrain with the help of small human guidance at the start of the retraining. To the best of our knowledge, this is the first paper to propose a digital twin with human-in-the-loop for training on-the-fly of RL for robot navigation.

We test the performance of our proposed model in both simulation and real-world environments with both static and dynamic obstacles. In summary, the major contributions of this paper are as follows:
\begin{itemize}
    \item We create a digital twin that continuously models the real-world environment in simulation to bridge the gap between sim-to-real transfer.
    \item We formulate a retraining mechanism that allows the physical twin to stop receiving controls when faced with a difficult scenario while the virtual twin explores and finds a solution.
    \item We built an assistive teleoperation human-in-the-loop with a keyboard which allows humans to directly send velocity commands to the virtual robot in real time during retraining. This allows the model to adapt in real-time as it receives feedback from human operators and accelerates the learning process.
\end{itemize}

The remainder of the paper is organized as follows. We present related work in Section \ref{sec:literature}. In Section \ref{sec:preliminaries}, we discussed the preliminary theories used in the research. Section \ref{sec:methodology} discusses the proposed methodology. Experimental analysis is described in \ref{sec:experiment}. Section \ref{sec:result} discusses the results and analysis. Finally, we conclude our major findings in Section \ref{sec:conclusion}.

\section{Related Work}
\label{sec:literature}
\subsection{Reinforcement Learning Approaches}
There has been substantial research on applying RL to UGV navigation. Mirowski et al. \cite{mirowski2018learning} developed an RL model that learns to navigate its environment from raw sensory inputs without relying on pre-built maps. In the work of \cite{olayemi2023impact}, they developed a Twin-Delayed Deep Deterministic policy gradient (TD3) network to investigate the impact of the sensor's field of views in the control of UGV navigation. Zhang et al. \cite{zhang2019reinforcement} demonstrated the implementation of a continuous control task using deep reinforcement learning for autonomous parking of UGVs.  Their work showcased the potential of DRL for real-time decision-making in complex environments, handling continuous action spaces effectively.  

\subsection{Human Input Reinforcement Learning Approaches}
RL is a promising approach for the UGV navigation system, but yet to reach, surpass or replace human driver's intelligence in handling unfamiliar situations. This motivates the introduction of human intelligence or expertise in reinforcement learning. A common way is incorporating human dataset trajectories using sequences of state-action pairs generated by humans performing a task. Hester et al. \cite{article} modified the loss function of Deep Q-learning to leverage a small dataset of human demonstrations to accelerate the learning process. They also introduced a modified replay buffer that combines the human expert demonstrations with the self-generated data by sampling a proportional amount of each type, ensuring that the sample data never overwrites the expert data. Vecerik et al. \cite{vecerik2018leveraging} developed a Deep Deterministic Policy Gradient (DDPG) from expert demonstration. They populate a replay buffer with demonstrations from human experts before training and self-interaction with the environment during training. The sampling ratio between the actions is tuned using a prioritized experience replay buffer. Though incorporating human dataset trajectories in RL helps with accelerated learning and improved performance, this method is faced with challenges such as handling covariate shifts and ensuring the quality of human demonstration.

In a bid to solve the problem with human datasets, researchers introduced real time human demonstration during training. Arakawa et al. \cite{Arakawa2018DQNTAMERHR} developed the DQN-TAMER algorithm that enables the human expert to observe the training agent's actions and continuously give an immediate reward as feedback. Wu et al. \cite{wu2023toward} developed human-guidance DRL by modifying the actor-critic policy and value networks to enable the transfer of control between humans and the automated system during training. However, due to the difference between its performance in simulation and real life, in \cite{wu2023human}, they introduce the ability of human agents to take over control of the agent in real time after deployment. Their approach allows humans to control the agent's action as deemed necessary, improving safety at runtime. In a similar work, Luo et al. \cite{luo2023human} proposed a Q value-dependant policy that allows the agent to implement selective human experts action provided at the early stage of the training based on the difference in Q value. These methods have also shown good progress in autonomous navigation, however, we cannot overlook the challenges that come with it such as requiring continuous human observation during the training process for feedback, the quality of human demonstrations and the learning algorithm's capability to generalize from human-guided experiences.

\section{Preliminaries}
\label{sec:preliminaries}
In this section, we discuss about Markov Decision Process (MDP) which serves as the mathematical framework used to model decision-making problems in situations where outcomes are partly random and partly under the control of a decision-maker. Next, we discuss TD3 \cite{9645287} and TD3 digital twin retraining \cite{olayemi2024twin} which we use as representative algorithms and baseline models for evaluating our methodology in this article. 

\subsection{Markov Decision Process}
Reinforcement Learning (RL) is a powerful approach to solving Markov Decision Processes (MDPs) when the model of the environment is not known a priori. RL within the context of an MDP is represented as \(S, A, R, P, \gamma\), where \(S\) denotes the set of all possible states \(s \in S\), \(a\) the set of all possible actions \(a \in A\), \(R\) is the immediate reward received after transitioning from one state to another due to an action, \(P\) is the probability of transitioning from one state to another given an action and \(\gamma \in [0,1]\) is the discount factor that defines the action to take in each state. The RL algorithms aim to find the optimal policy \(\pi^*\) without prior knowledge of the transition probabilities and reward functions. The agent explores the environment, collects experiences, and improves its policy based on the feedback received. The expected return is estimated using the state-value Bellman equation: 
\begin{equation}
    V(s) = \sum_{a \in A} \pi(a|s) \sum_{s' \in S} P(s'|s, a) \left[ R(s, a, s') + \gamma V(s') \right]
\end{equation}
representing the expected return starting from state \(s\) and following policy \(\pi\) or the action-value Bellman equation: 
\begin{equation}
    Q(s, a) = \sum_{s' \in S} P(s'|s, a) \left[ R(s, a, s') + \gamma \sum_{a' \in A} \pi(a'|s') Q(s', a') \right]
\end{equation}
representing the expected return starting from state \(s\), taking action \(a\) and then following policy \(\pi\).

\subsection{TD3 Algorithm}
TD3 is an advanced RL algorithm designed to address the limitations and improve the performance of Deep Deterministic Policy Gradient (DDPG). TD3 is an off-policy algorithm that extends DDPG by introducing several key enhancements that stabilize training and reduce overestimation bias. Like DDPG, TD3 uses an actor-network (policy) \(\pi_\theta\) which selects actions, but two critic-networks \(Q_{\phi1}\) and \(Q_{\phi2}\) which evaluates those actions by estimating the Q-values. The target Q-value is computed by taking the minimum value predicted by the two critics.

To prevent the actor-network from being updated too frequently, the actor is updated once every two or more critic updates. This helps in stabilizing the learning process as the policy is updated using more accurate Q-values. To mitigate the issue of overfitting to narrow peaks in the value function, TD3 adds noise to the target actions. This technique, known as target policy smoothing, helps to create smoother target value estimates by applying a small random perturbation to the actions used in target Q-value computation. The general procedure of TD3 is presented in Algorithm \ref{alg:TD3}.
\begin{algorithm}[H]
\caption{Twin Delayed Deep Deterministic Policy Gradient (TD3)}
\label{alg:TD3}
\begin{algorithmic}[1]
\STATE Initialize critic networks $Q_{\theta_1}(s, a)$ and $Q_{\theta_2}(s, a)$ with random parameters $\theta_1$ and $\theta_2$
\STATE Initialize actor network $\pi_{\phi}(s)$ with random parameters $\phi$
\STATE Initialize target networks $Q_{\theta_1'}(s, a)$, $Q_{\theta_2'}(s, a)$, and $\pi_{\phi'}(s)$ with $\theta_1' \leftarrow \theta_1$, $\theta_2' \leftarrow \theta_2$, $\phi' \leftarrow \phi$
\STATE Initialize replay buffer $\mathcal{D}$
\FOR{each episode}
    \FOR{each step $t$ in the episode}
        \STATE Select action $a_t = \pi_{\phi}(s_t) + \epsilon$, with $\epsilon \sim \mathcal{N}(0, \sigma)$
        \STATE Execute action $a_t$ and observe reward $r_{t+1}$ and next state $s_{t+1}$
        \STATE Store transition $(s_t, a_t, r_{t+1}, s_{t+1})$ in replay buffer $\mathcal{D}$
        \IF{it's time to update}
            \STATE Sample mini-batch of $N$ transitions $(s_i, a_i, r_i, s_{i+1})$ from $\mathcal{D}$
            \STATE Compute target actions with noise: $a'_{i+1} = \pi_{\phi'}(s_{i+1}) + \epsilon'$, with $\epsilon' \sim \text{clip}(\mathcal{N}(0, \sigma'), -c, c)$
            \STATE Compute target Q-value:
            \[
            y_i = r_i + \gamma \min(Q_{\theta_1'}(s_{i+1}, a'_{i+1}), Q_{\theta_2'}(s_{i+1}, a'_{i+1}))
            \]
            \STATE Update critic networks by minimizing the loss:
            \[
            L(\theta_j) = \frac{1}{N} \sum_i \left(Q_{\theta_j}(s_i, a_i) - y_i\right)^2 \quad \text{for } j = 1, 2
            \]
            \IF{it's time to update the policy}
                \STATE Update policy by maximizing the Q-value:
                \[
                \nabla_{\phi} J(\phi) = \frac{1}{N} \sum_i \nabla_a Q_{\theta_1}(s_i, a) \bigg|_{a=\pi_{\phi}(s_i)} \nabla_{\phi} \pi_{\phi}(s_i)
                \]
                \STATE Update target networks with soft updates:
                \[
                \theta_j' \leftarrow \tau \theta_j + (1 - \tau) \theta_j' \quad \text{and} \quad \phi' \leftarrow \tau \phi + (1 - \tau) \phi'
                \]
            \ENDIF
        \ENDIF
    \ENDFOR
\ENDFOR
\end{algorithmic}
\end{algorithm}

\subsection{TD3 Digital Twin Retraining}
The Digital Twin (DT) technology is a crucial and rapidly developing tool for digital transformation and smart enhancements \cite{tao2022digital,singh2021digital,bottjer2023review}. It allows the virtual representation of a physical object, system, or process. The digital model mirrors the characteristics and behaviors of its real-world counterpart, allowing real-time monitoring, simulation, and optimization. In other to bridge the gap between sim-to-real transfer of DRL models, in our previous work \cite{olayemi2024twin}, we mirrored the behaviour and environment of a physical UGV in simulation. Once the digital twin is created, the simulated UGV is controlled using the trained model verifying the path is collision free then send same control to the physical UGV. During navigation, if there is a difficult situation the trained model could not solve, it enters into a retraining mode in simulation using the real time environment state of the real world mirrored. In this paper, the idea of TD3 digital twin retraining is also adopted to incorporate human expert input during the retraining. In the retraining process, a human expert uses teleoperation to control a few steps of the UGV and then leaves the control to the automated system to overcome the difficult task.

\section{Methodology}
\label{sec:methodology}
\begin{figure}%[!t]
\centering
\includegraphics[width=3.5in]{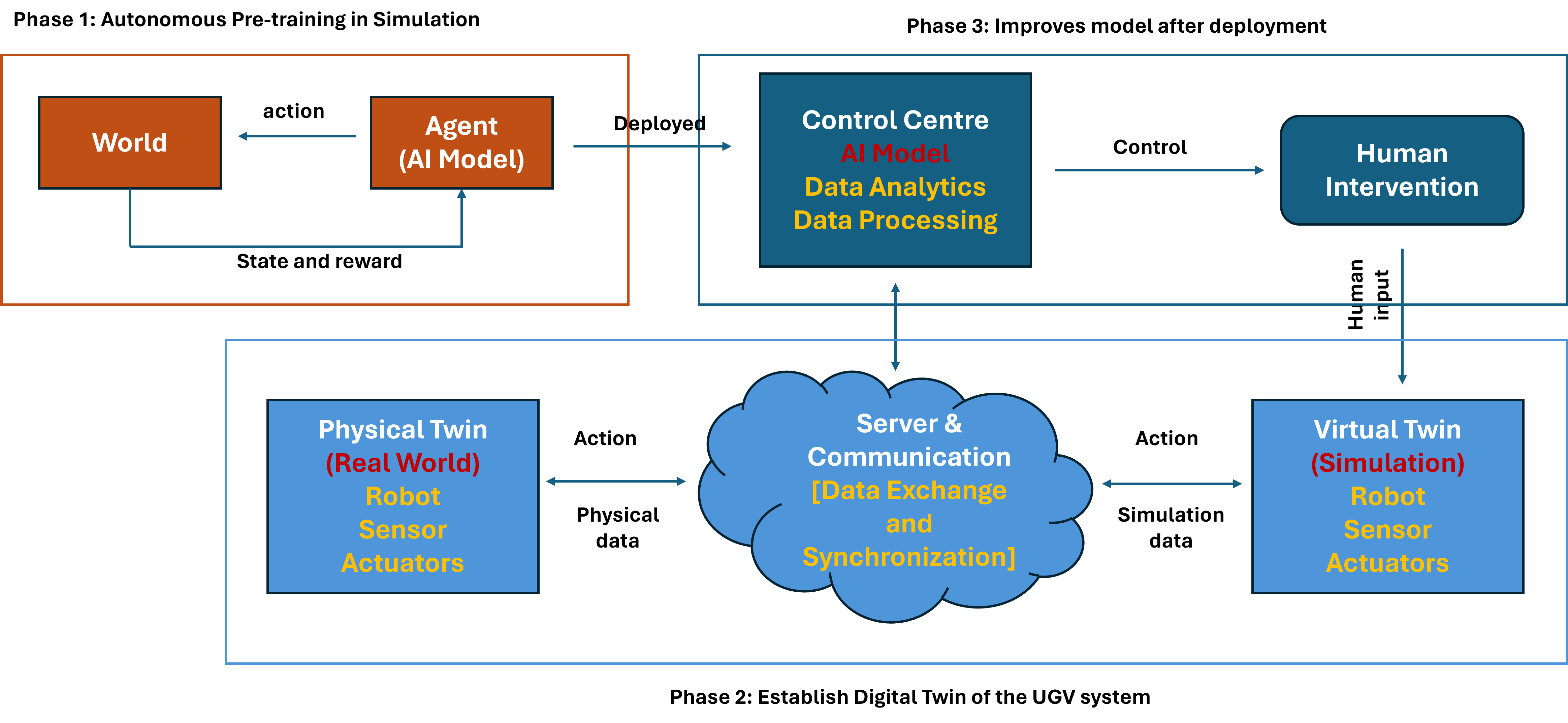}
\caption{We present the overall framework of pur proposed methodology. In phase 1, we pre-trained the TD3 agent in simulation storing experiences of simulation in both replay and priority buffers. The deployment integrates both phases 2 and 3. Phase 2 presents the creation of the digital twin and phase 3 controls the operation of the whole system switching between the AI model or human expert guidaiance.  }
\label{fig:model}
\end{figure}

In this section, we propose a human-in-the-loop digital twin retraining system to address (1) sim-to-real transfer issue faced with RL models, and (2) improve the performance of the RL model after deployment by incorporating human expert input during retraining. The architecture of our proposed model is presented in Figure \ref{fig:model}, which is divided into three phases. The first phase is the pre-training phase, the second phase is the creation of digital twin for deployment and the third phase is the retraining phase with human expert input.
\subsection{Pre-Trained Model}
\label{subsec:td3}
An RL model is pre-trained in simulation by continuously interacting with its environment, collecting experiences, updating the network weights, and periodically evaluating the performance of the network as presented in Algorithm \ref{alg:pretrain_td3}. The actor-network is a neural network that outputs the actions to be taken given the current state. The state \(s=[z,d,\phi,a_1,a_2]\) includes the information from the LiDAR sensor \(z\), the distance of the robot to it's goal \(d\), the orientation of the robot \(\phi\) and the current actions taken by the robot \(a_1,a_2\). The structure of the network is designed to consist of three fully connected layers with ReLU activations, except for the final layer, which uses a Tanh activation to ensure the action outputs are bounded between -1 and 1.
\begin{equation}
    a = \text{tanh}(W_3 \cdot \text{ReLU}(W_2 \cdot \text{ReLU}(W_1 \cdot s + b_1) + b_2) + b_3)
\end{equation}
where \(s\) is the state vector representing the environment state, \(W_1,W_2,W_3\) represents the weight matrices for the three fully connected layers and \(b_1, b_2, b_3\) represents the bias vectors for the three layers. We define the actor loss as the negative Q-value predicted by the critic network.
\begin{equation}
    \text{Actor Loss} = -Q_1(s, \text{Actor}(s))
\end{equation}
The critic network evaluates the Q-value of taking a certain action in a given state. TD3 uses two critic networks (\(Q_1\) and \(Q_2\)) to reduce overestimation bias. Each critic network consists of three fully connected layers, with the second layer split into two parts to separately process the state and action inputs before combining them through additional layers to produce the Q-value.
\begin{align}
s_1 &= \text{ReLU}(W_{1\_1} \cdot s + b_{1\_1}) \\
s_2 &= \text{ReLU}(W_{2\_1} \cdot s_1 + b_{2\_1}) \\
Q_{1\_s} &= W_{3\_1} \cdot s_2 + b_{3\_1} \\
a_1 &= \text{ReLU}(W_{1\_2} \cdot a + b_{1\_2}) \\
a_2 &= \text{ReLU}(W_{2\_2} \cdot a_1 + b_{2\_2}) \\
Q_{1\_a} &= W_{3\_2} \cdot a_2 + b_{3\_2} \\
Q(s, a) &= Q_1(s, a) = Q_2(s, a) = Q_{1\_s} + Q_{1\_a}
\end{align}
where \(s\) is the state vector, \(a\) is the action vector representing the action taken by the agent, \(W_{1\_1}, W_{2\_1}, W_{3\_1}\) represents the weight matrices for the state pathway, \(W_{1\_2}, W_{2\_2}, W_{3\_2}\) represents the matrices for the action pathway, \(b_{1\_1}, b_{2\_1}, b_{3\_1}\) represent the bias vectors for the layers in the state pathway, and \(b_{1\_2}, b_{2\_2}, b_{3\_2}\) represent the bias vectors for the layers in the action pathway. The critic loss is defined as mean squared error between the predicted Q-values and the target Q-values.
\begin{equation}
    \text{Critic Loss} = \frac{1}{N} \sum_{i=1}^{N} \left( Q_1(s_i, a_i) - y_i \right)^2
\end{equation}
where \(N\) represents the batch size, \((s_i,a_i)\) are the state state-action pairs sampled from the buffer, and \(y_i\) represents the target Q-values.
 
Traditionally, the TD3 network experience replay is implemented using a replay buffer, which stores past experiences tuples (state, action, reward, next state, done) in a circular buffer with new experiences replacing older ones once the buffer is full. In our work, we introduce a priority replay buffer which stores prioritizes experiences based on their Temporal Difference error (TD-error). This helps to train the network more efficiently by focusing on experiences that are more informative.
\begin{algorithm}[H]
\caption{Pre-Trained TD3 Algorithm}
\label{alg:pretrain_td3}
\begin{algorithmic}[1]
\STATE Initialize actor and critic networks with random weights
\STATE Initialize target networks with the same weights as the networks
\STATE Initialize replay buffer and priority buffer
\STATE Initialize exploration noise parameters
\STATE Set training hyperparameters: batch size, discount factor, soft target update rate, etc.
\FOR{each training episode}
    \STATE Reset environment and initialize state
    \FOR{each timestep in episode}
        \STATE Select action using actor network with added exploration noise
        \STATE Execute action, observe reward and next state
        \STATE Store transition in replay buffer and priority buffer
        \STATE Sample batch from buffers and perform gradient descent on actor and critic networks
        \STATE Update target networks using soft update strategy
    \ENDFOR
\ENDFOR
\end{algorithmic}
\end{algorithm}

\subsection{Digital Twin}
In this phase, a digital twin of the physical robot is created to enable seamless interaction between the physical and virtual environments. The physical twin is responsible for managing the robot's navigation and processing sensor data in real-world scenarios. At each timestep \(t\), it collects data from physical sensors \(z_t\) to detect obstacles and their positions, \(o_t = f(z_t)\), where \(f\) is a function representing the obstacle detection algorithm and then transmits this information in real time to the virtual twin.

The virtual twin utilizes this real-time sensor data \(o_t\) to accurately replicate the physical environment within a simulation. This capability allows for the exploration of various "what-if" scenarios without any risk to the physical robot, facilitating the testing and optimization of different navigation strategies under simulated conditions.

Once the physical environment \(s_t\) is successfully mirrored in simulation \(\hat{\mathbf{s}}_t\), a trained model is deployed to control \(a_t\) the navigation of the virtual robot \(\hat{\mathbf{s}}_{t+1} = \hat{g}(\hat{\mathbf{s}}_t, \mathbf{a}_t)\) where \(\hat{g}\) is the transistion function in the virtual environment. This control strategy is applied to the physical robot in real time \(\mathbf{s}_{t+1} = g(\mathbf{s}_t, \mathbf{a}_t)\) where \(g\) is the transition function in the real environment, provided no immediate risks are detected. The virtual twin serves as a crucial testbed for validating navigation strategies, ensuring they are both safe and effective before real-world implementation.

In instances where an impending risk is detected or if the pre-trained model under-performs, the real-time control transfer to the physical robot is halted. Instead, the system initiates retraining of the TD3 model using the current environmental state. This retraining process is critical for adapting the model to new scenarios, enhancing its performance, and mitigating potential risks. Upon completion of retraining, the updated TD3 model is reloaded, and navigation resumes within the digital twin environment. The overall structure of the digital twins is presented in Algorithm \ref{alg:digital_twin}
\begin{algorithm}[H]
\caption{Digital Twin Process for Robotic Navigation}
\label{alg:digital_twin}
\begin{algorithmic}[1]
\STATE \textbf{Initialize:} Set up physical and virtual robot environments.
\STATE Initialize physical robot with real-time sensors.
\STATE Initialize virtual twin with simulation capabilities.
\STATE Initialize TD3 model with actor $\pi_\phi$ and critics $Q_{\theta_1}$, $Q_{\theta_2}$

\FOR{each operational cycle}
    \STATE Reset environment and initialize states $\mathbf{s}_t$ and $\hat{\mathbf{s}}_t$
    \FOR{each timestep $t$}
        \STATE \textbf{Data Collection:}
        \STATE Collect and process sensor data $\mathbf{z}_t$ to detect obstacles $\mathbf{o}_t$
        \STATE Transmit processed data $\mathbf{o}_t$ to virtual twin
        
        \STATE \textbf{Virtual Environment Simulation:}
        \STATE Simulate virtual state transition $\hat{\mathbf{s}}_{t+1} = \hat{g}(\hat{\mathbf{s}}_t, \mathbf{a}_t)$
        \STATE Optimize policy $\pi$ to maximize expected cumulative reward
        \STATE Generate control action $\mathbf{a}_t = \pi(\hat{\mathbf{s}}_t)$
        
        \STATE \textbf{Control Deployment:}
        \IF{no risk detected}
            \STATE Apply $\mathbf{a}_t$ to physical robot $\mathbf{s}_{t+1} = g(\mathbf{s}_t, \mathbf{a}_t)$
        \ELSE
            \STATE Halt control transfer and initiate retraining
        \ENDIF
    \ENDFOR

    \STATE Simulate various navigation scenarios and strategies.
    \IF{no risks are detected in simulation}
        \STATE Apply tested strategies to control the physical robot.
    \ELSE
        \STATE Trigger retraining and strategy adjustment protocols.
    \ENDIF
\ENDFOR
\STATE \textbf{Model Retraining:}
\IF{risk detected or suboptimal performance}
    \STATE Sample batches from replay and priority buffers
    \STATE Update critics $Q_{\theta_1}$ and $Q_{\theta_2}$
    \STATE Update actor $\pi_\phi$ and target networks
    \STATE Transfer control back to the physical robot after retraining
\ENDIF
\end{algorithmic}
\end{algorithm}

\subsection{Human-in-the-Loop TD3}
A key use case for this phase occurs when the existing navigation model proves inadequate due to unfamiliar environments or novel tasks, necessitating immediate, on-the-fly retraining of the TD3 model using the current environmental state. The retraining leverages past experiences stored in the replay and priority buffers to adapt the model to new scenarios and improve its performance. During this retraining phase, human experts may intervene by using keyboard inputs to manually control the initial navigation steps of the robot within the simulation. This intervention ensures that the model does not excessively mimic human input, which could hinder its ability to generalize beyond the specific scenarios encountered during retraining sessions. Once the robot successfully navigates to its target in the simulation, the retraining concludes, and the robot is returned to its original position within the digital twin to continue its navigation. Algorithm \ref{alg:human_in_the_loop} gives an overview of the human-in-the-loop process.

This dual-phase approach combines automated learning with manual oversight to maintain robustness and adaptability in dynamic environments. It balances algorithm-driven decision-making with human expertise to refine the robot's operational capabilities continuously.
\begin{algorithm}[H]
\caption{Human Intervention in Retraining of Robotic Navigation}
\label{alg:human_in_the_loop}
\begin{algorithmic}[1]
\STATE \textbf{Initialize:} Set up interfaces for human-robot interaction.
\STATE Connect teleoperation interface with virtual twin simulation and control systems.
\STATE Prepare systems for emergency manual control and guidance.

\STATE \textbf{Retraining Loop:}
\WHILE{retraining is active}
    \STATE Monitor for impending risks or suboptimal model performance.
    \IF{risk detected or suboptimal performance}
        \STATE Enable human control mode.
        \STATE Humans guide the robot manually using the teleoperation interface.
        \STATE Collect data from human-controlled sessions for training.
        \STATE Use human input data to adjust and refine the model.
        \STATE Gradually reduce human input as model performance improves.
        \STATE Resume full autonomous control in both virtual and physical environments.
    \ENDIF
\ENDWHILE
\STATE Ensure the model does not overly focus on human-like behaviour.
\end{algorithmic}
\end{algorithm}
\section{Experimental Design}
\label{sec:experiment}
In this section, we discuss the experimental setup for our proposed methodology to verify its performance in UGV navigation tasks. Hence, we design our experiment to establish the following: (1) to validate that the design of a digital twin for navigation task bridges the gap between sim-to-real transfer, (2) to validate the ability of RL model to self-improve in real time after the deployment, (3) to validate that human expert guidance can create a more robust and adaptive navigation model. To conduct a comparative study compare our methodology with TD3 \cite{9645287} and TD3DT \cite{olayemi2024twin} algorithms using the same parameter settings given in Table \ref{tab:hyperparameters}.
\begin{table}[!t]
\caption{TD3 Model Hyperparameters\label{tab:hyperparameters}}
\centering
\begin{tabular}{|c|c|}
\hline
Hyperparameter & Value \\
\hline
Device & `cuda` or `cpu` \\
\hline
Max Steps per Episode & 500 \\
\hline
Initial Exploration Noise & 1 \\
\hline
Exploration Decay Steps & 500,000 \\
\hline
Learning Rate (Alpha) & 0.6 \\
\hline
Minimum Exploration Noise & 0.1 \\
\hline
Batch Size & 40 \\
\hline
Discount Factor & 0.99999 \\
\hline
Tau (Soft Update Rate) & 0.005 \\
\hline
Policy Noise & 0.2 \\
\hline
Noise Clip & 0.5 \\
\hline
Policy Update Frequency & 2 \\
\hline
Replay Buffer Size & 1e6 \\
\hline
Priority Buffer Size & 1e6 \\
\hline
\end{tabular}
\end{table}
\subsection{Experimental Setup}
The computer used for the experimental simulation is equipped with an Intel Core i7-6800 CPU desktop computer with 32 GB RAM and an NVIDIA GTX 1050 graphics card. The simulation software used is Gazebo as it is supported by both Ubuntu 20.04 operating system and ROS Noetic. To ensure comparison fairness, our proposed methodology and other baseline algorithms for comparison are implemented using the same computer and implemented using the PyTorch framework in Python. Also, we ensure that the same environment complexity is used for all the algorithms.

\begin{figure}%[!t]
\centering
\includegraphics[width=2.5in]{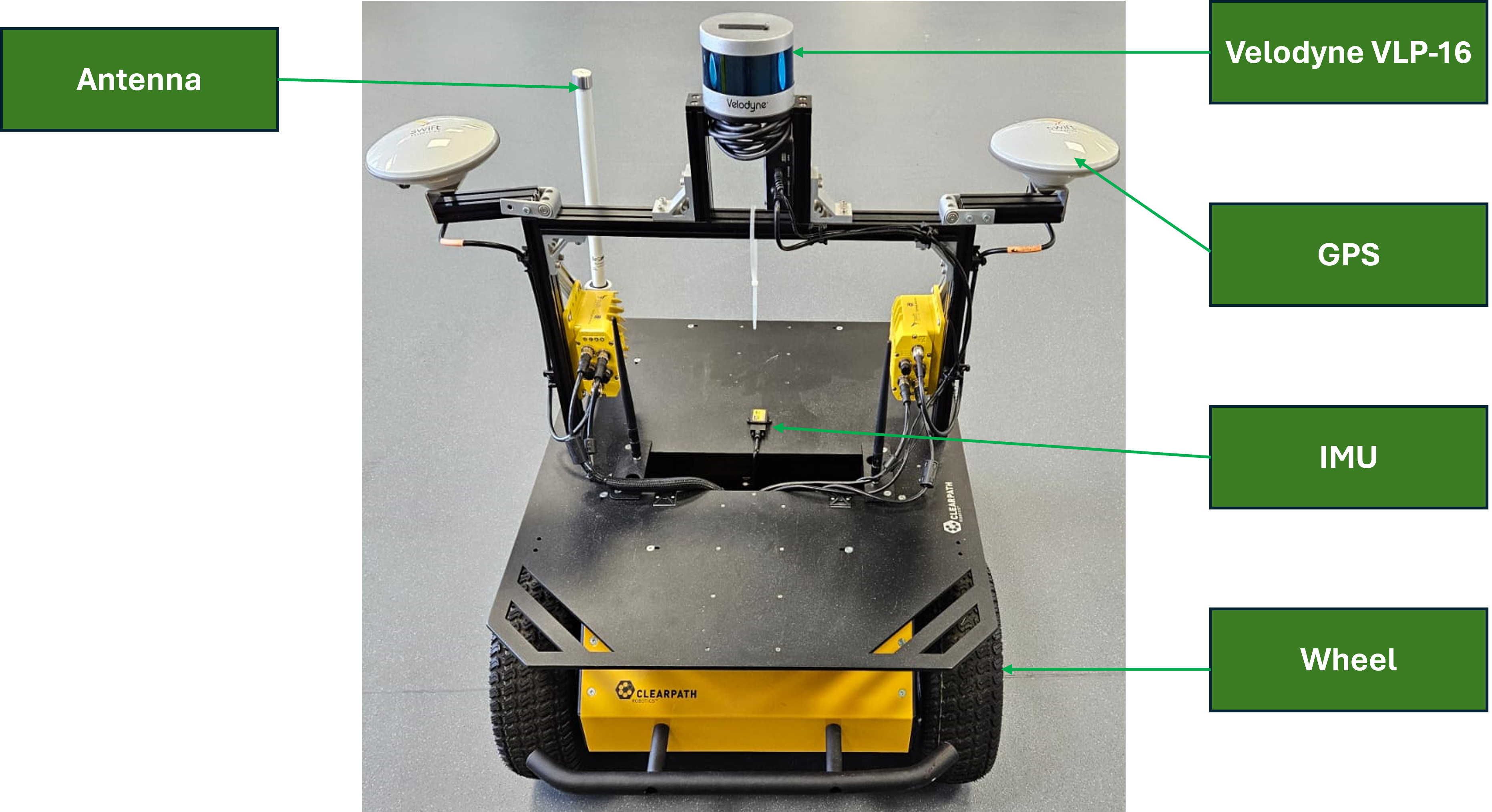}
\caption{Husky A200 UGV is a differential-drive implemented on Ubuntu 20.04 operating system and supported by Robot Operating System (ROS).}
\label{fig:husky}
\end{figure}

Both in simulation and real-world experiments we used Husky A200 UGV (see Figure \ref{fig:husky}) developed by Clearpath Robotics, Inc., Ontario, Canada. It is a non-holonomic differential-drive vehicle supporting a maximum of \(1m/s\) linear and angular movements. It has a mini ITX computer, Global Positioning System (GPS) and open Inertial Measurement Unit (IMU). A VLP-16 3D LiDAr sensor is mounted on it for environmental perception. The UGV has Ethernet ports, an antenna and a base station for remote data exchange and control support. 

\subsection{Human Expert Intervention}
Following deployment, we develop a mode for human expert intervention that allows for smooth integration and withdrawal of human expertise within the RL process.  With the help of real-time simulation of the real world through digital twin, human experts can have the same perception information as the real robot. Human expert gives control information using keyboard teleoperation. We designed nine different types of keypress controls, including forward-right, backwards-right, forward-left, backwards-left, backward, forward, turn-right, turn-left, and stop for the human expert use. With this teleoperation, a human expert can decide at any time to give commands to the robot or not. The velocity information and the associated keypress are described in Table \ref{tab:teleoperation}.

\textit{Remark 1:} In this paper, we use a keyboard teleoperation to get the inputs from human operator. However, similar teleoperation devices such as haptic, virtual reality (VR), 3D mouse, etc, can be used to capture the human control input better. 

\begin{table}[!t]
\caption{Keyboard Presskey Velocities for Teleoperation Mode\label{tab:teleoperation}}
\centering
\begin{tabular}{|c|c|c|}
\hline
Key & Linear Velocity & Angular Velocity \\
\hline
w (forward-right) & 0.5 & 0.5 \\
\hline
z (Backward-right) & -0.5 & 0.5 \\
\hline
a (forward-left) & 0.5 & -0.5 \\
\hline
d (backward-left) & -0.5 & -0.5 \\
\hline
l (left) & 0.0 & -0.5 \\
\hline
r (right) & 0.0 & 0.5 \\
\hline
f (forward) & 0.5 & 0.0 \\
\hline
b (backward) & -0.5 & 0.0 \\
\hline
s (Stop human input)& 0.0 & 0.0 \\
\hline
\end{tabular}
\end{table}

\subsection{Simulation Environment Complexity}
\begin{figure}%[!t]
\centering
\includegraphics[width=2.5in]{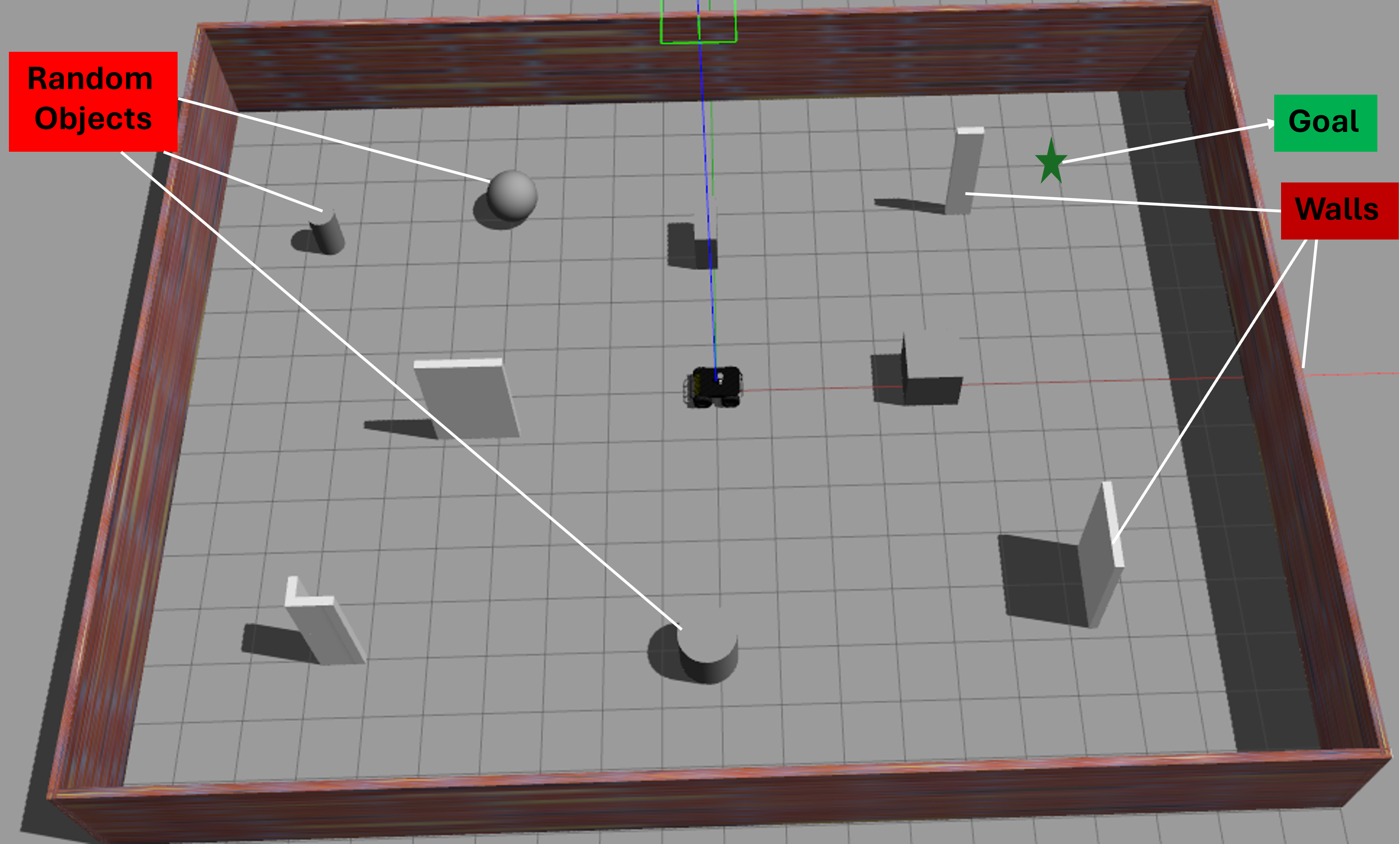}
\caption{At the start of each episode during training in the simulation, the UGV is randomly spawned at different locations maintaining a safe distance between both the goal(green star) and and the obstacles. The walls in the world are static throughout the training while other obstacles i.e. the box, cylinder and sphere are also randomly positioned at the start of a new episode.}
\label{fig:simulation}
\end{figure}

\begin{figure}%[!t]
\centering
\includegraphics[width=3.5in]{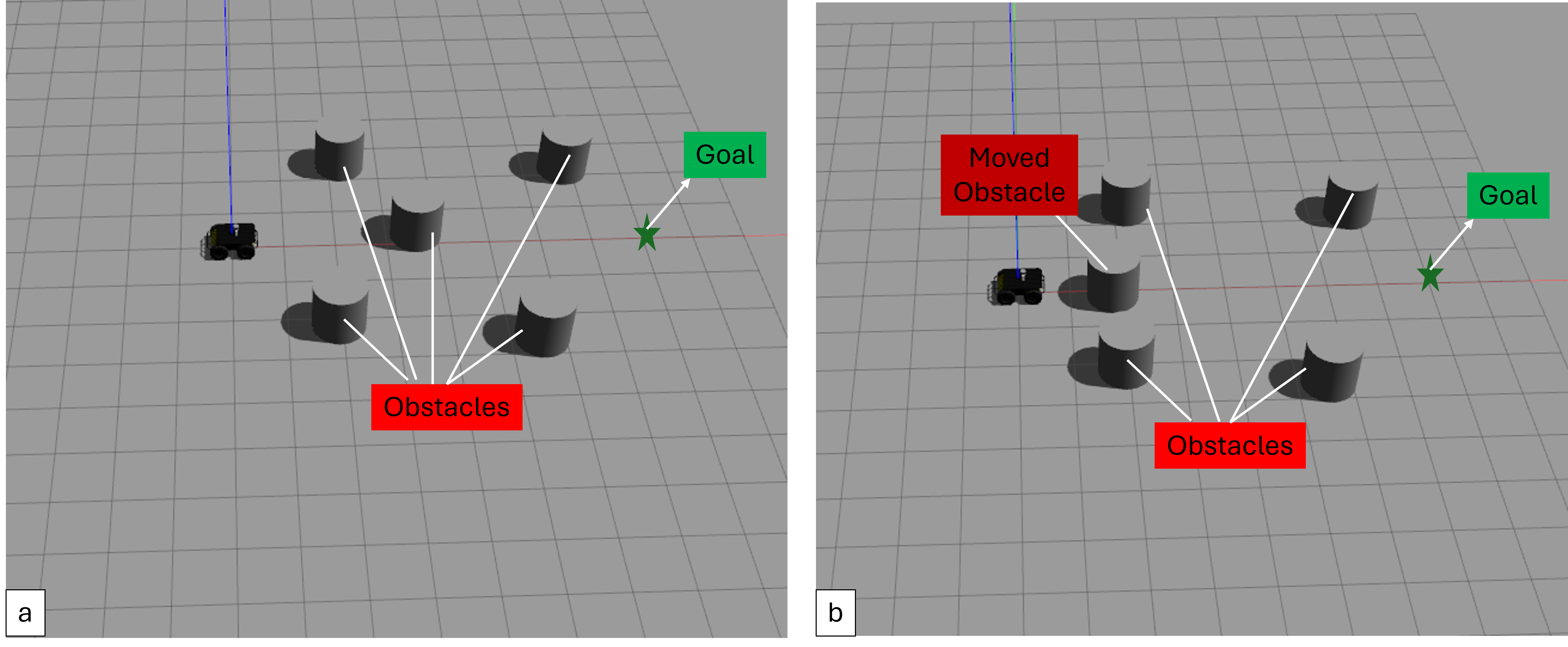}
\caption{Examples of environment scenes used for testing the model (a) In this environment scene, all objects are static with a minimum distance of over \(2 m\) between the robot and the obstacles. (b) In this environment, we moved one of the obstacles closer to the robot maintaining just \(1.1 m\) distance between the robot and the obstacle.}
\label{fig:simulation_test}
\end{figure}

To train the RL algorithms, we create an environment consisting of different sizes and shapes of static obstacles within a \(20 \times 15 m\) rectangle-shaped wall to limit the exploration space. At the start of each episode, the position of the robot is randomly spawned into the environment ensuring at least \(2 m\) distance to other obstacles and \(3 m\) distance to the goal. Also, some obstacles (such as the sphere, cylinder, and box shapes) are randomly placed in other to randomize the training process as shown in Figure \ref{fig:simulation}. During simulation testing, we adopt a different environment setup (see Figure \ref{fig:simulation_test}) from the one used during training to check the generalization of the models. 
\subsection{Real-World Environment Complexity}
\begin{figure}%[!t]
\centering
\includegraphics[width=3.5in]{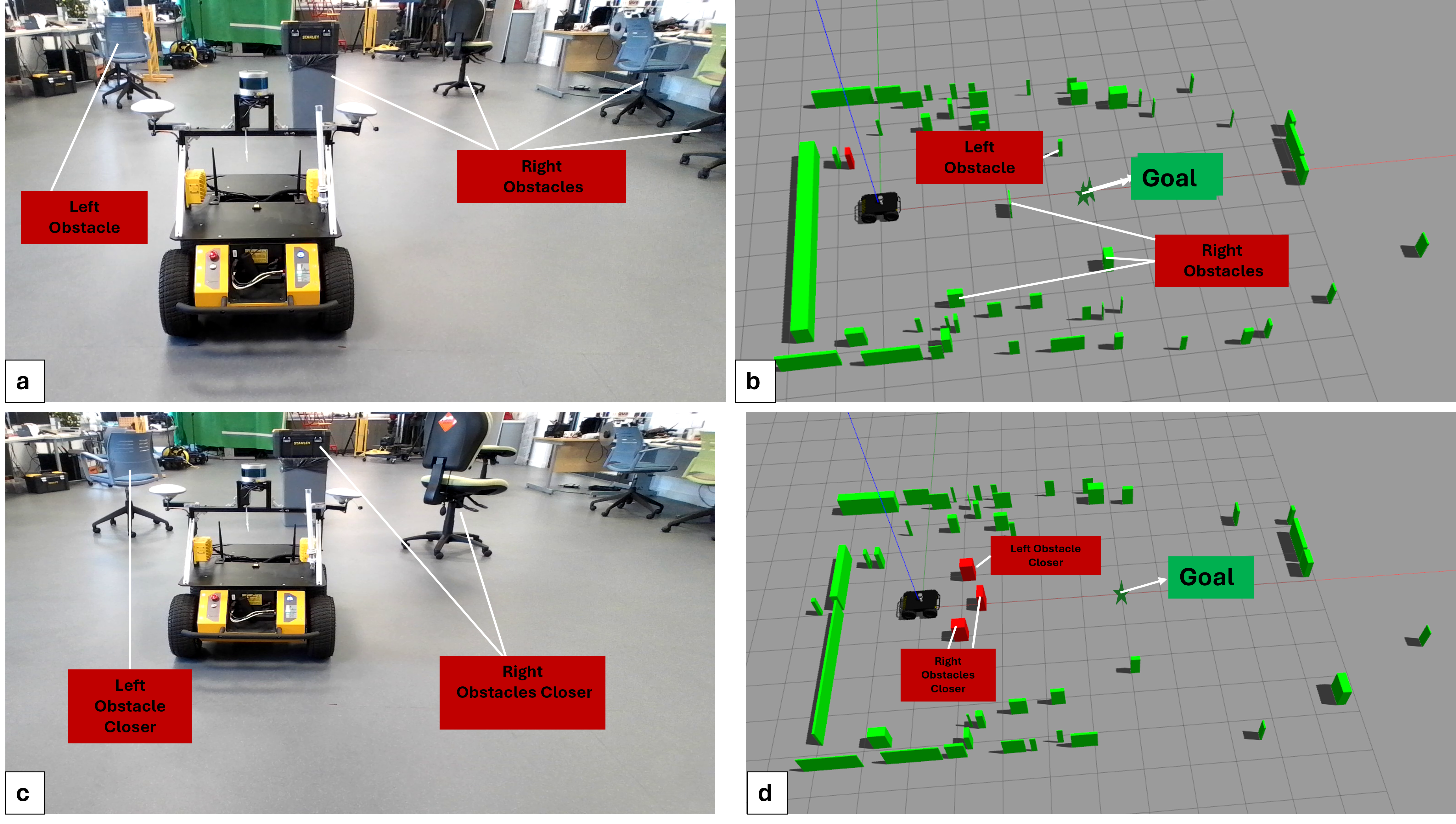}
\caption{This figure shows some of the setups of the environment used for training our model. The setup (a) consists of sparely distributed static objects. (b) is showing the twin of (a) in simulation. Similarly, (c) is another environment setup with densely distributed obstacles and (d) is its representative twin in simulation. The green boxes represent the objects in the physical world that are above \(1.5 m\) away from the robot while the red boxes are obstacles below \(1.5m\) way to the robot. This helps to easily track impending obstacles. }
\label{fig:realworld}
\end{figure}

In the real world, we experiment with our models and the baseline models on both static and dynamic obstacles as shown in Figure \ref{fig:realworld}. Dynamic obstacles include moving humans or sliding chairs. The UGV aims to navigate to its target point without colliding with any obstacles. We introduce dynamic obstacles during testing to investigate the ability of the model to self-improve.
\subsection{Evaluation Metrics}
This section describes the metrics used to evaluate the navigation performance in both simulation and real-world experiments. During simulation training, in each episode, we use the accumulated reward to determine the robot's learning rate, number of steps taken and travel time to evaluate the training performance. For the simulation testing, we use the success rate, which we define as the proportion of trials successfully completed, the collision rate measures the frequency of collisions during trials, indicating the safety and reliability of the navigation system, the timeout rate reflects the occurrences where the task was not completed within the allocated time, suggesting issues with efficiency or planning and lastly, the travel time measures the time taken to complete tasks, providing insight into the efficiency and speed of the system \cite{alhmiedat2023slam, kastner2023predicting, golchoubian2024uncertainty}. 
\section{Results and Analysis}
\label{sec:result}
In this section, we present the results of our developed methodology alongside the baseline TD3 and TD3DT algorithms during simulation training. Following this, we discuss the outcomes from testing the models in simulation and, finally, their performance after deployment on the real-world Husky A200 UGV. 
\subsection{Results in Simulation Training}
\begin{figure}%[!t]
\centering
\includegraphics[width=2.5in]{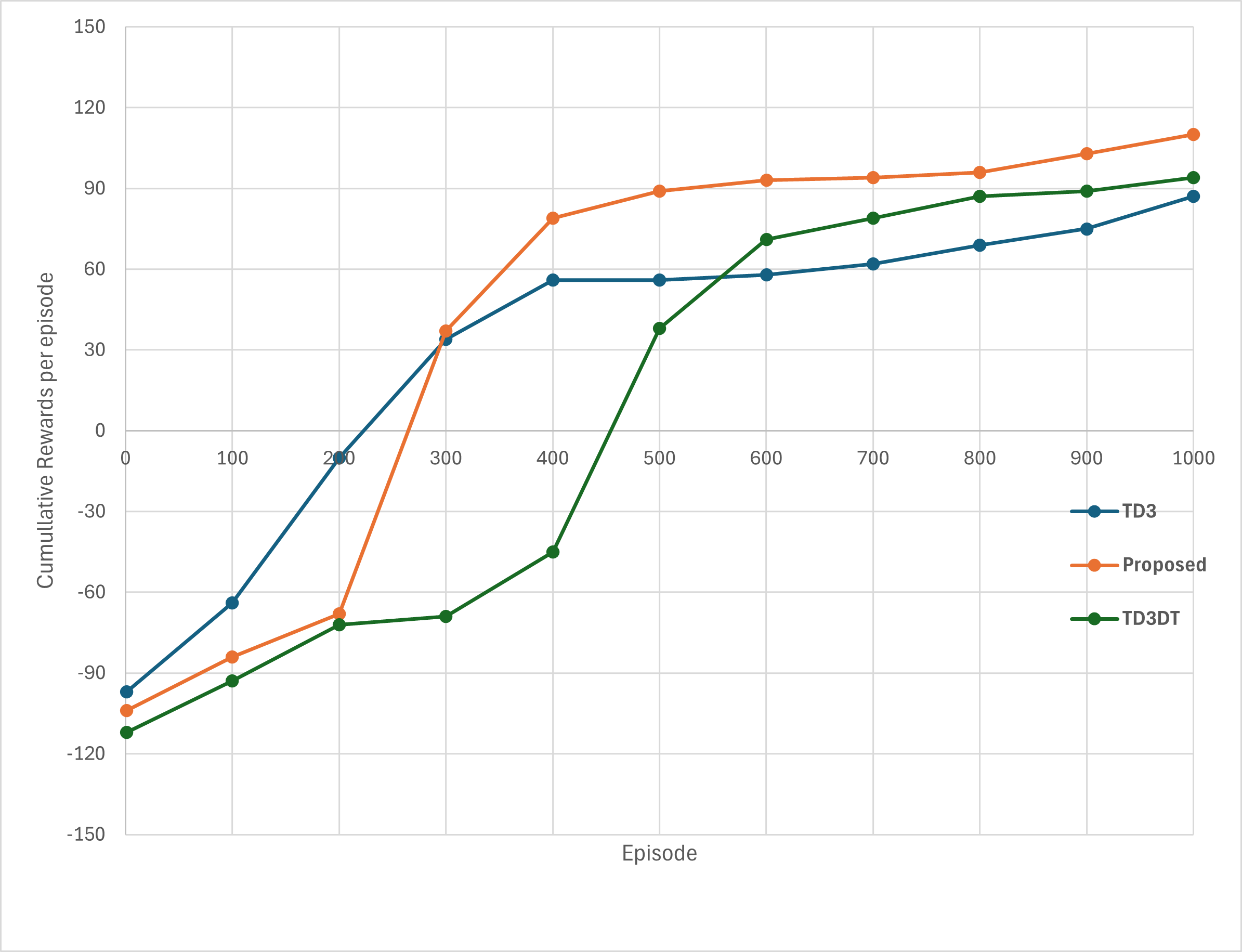}
\caption{Training rewards of our proposed methodology and other two baseline algorithms. }
\label{fig:train_result}
\end{figure}

The results presented in Figure \ref{fig:train_result} were derived from a comparative training session involving our innovative methodology and two advanced baseline algorithms, TD3 and TD3DT, each undergoing a rigorous 1000-episode training regime. Throughout this training period, we meticulously tracked the cumulative rewards at each episode’s time step. This granular data collection facilitated ongoing adjustments to the TD3 hyperparameters, optimizing our approach in real-time. A comparative analysis of episode rewards reveals that our methodology not only outperformed the baseline models in terms of reward accumulation but also demonstrated superior training efficiency. These findings underscore the effectiveness of our approach in leveraging advanced reinforcement learning techniques to achieve enhanced performance outcomes.
\subsection{Results in Simulation Testing}
\begin{figure*}[!t]
\centering
\subfloat[]{\includegraphics[width=2.5in]{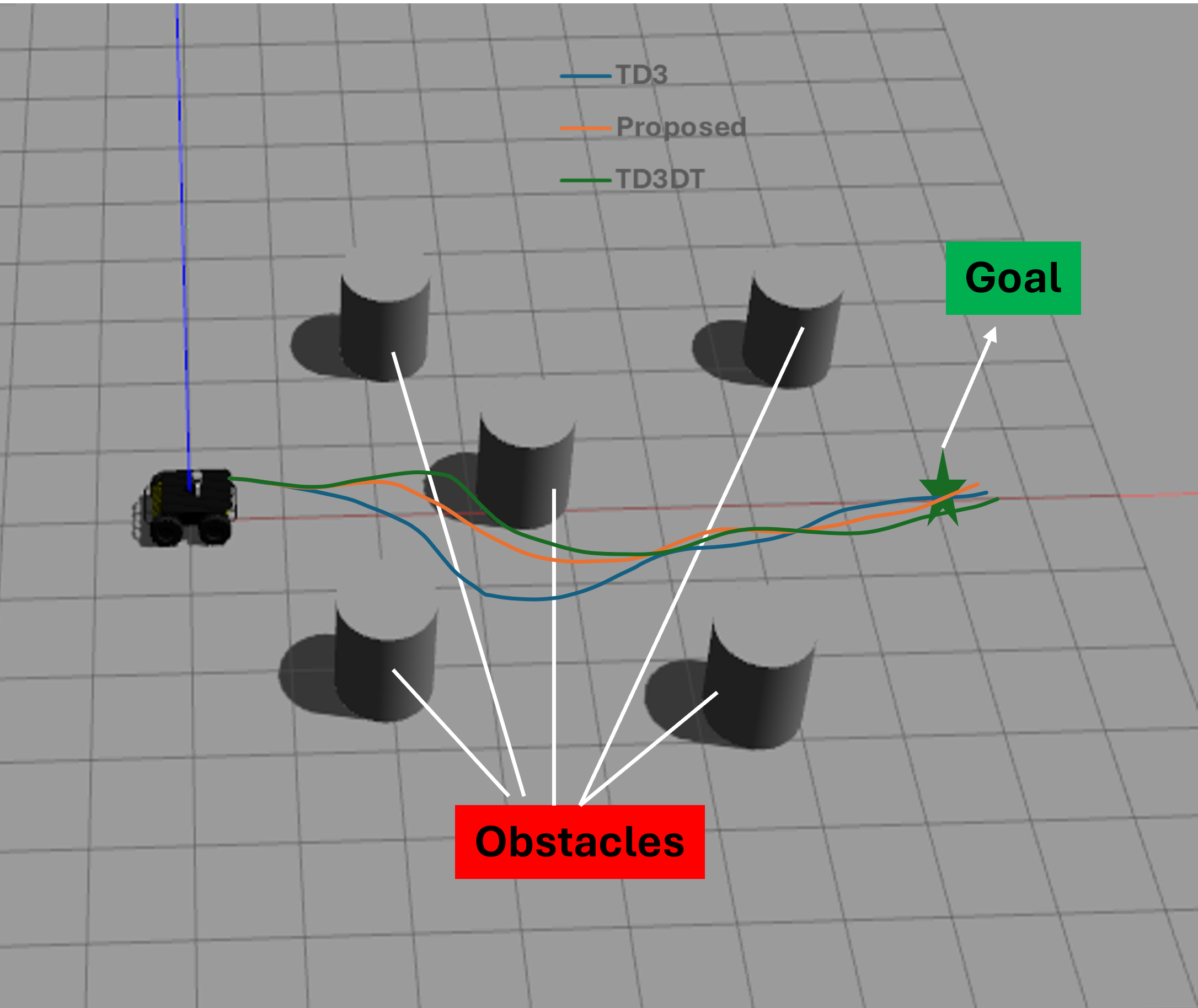}}%
\label{fig:simulation_test_result1}
\hfil
\subfloat[]{\includegraphics[width=2.5in]{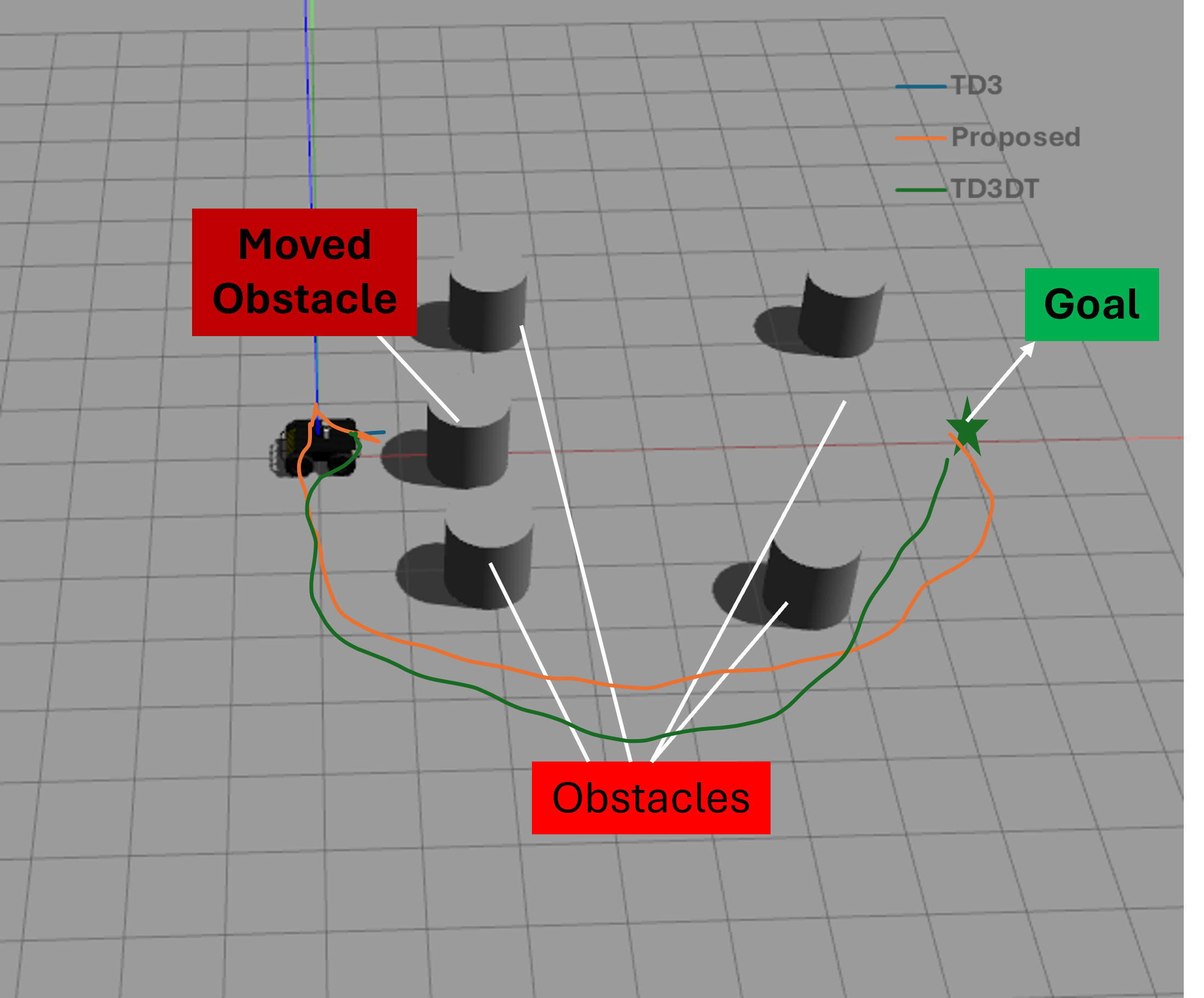}%
\label{fig:simulation_test_result2}}
\caption{Two different scenarios demonstrating the TD3 models. In case (a), both the baseline models and our models were able to navigate safely within the world. In case (b), both our model and TD3DT were able to reach their target. However, in both cases here, our proposed model shows a smoother and faster trajectory}.
\label{fig:simulation_test_result}
\end{figure*}

The outcomes of several simulation testing experiments are depicted in Figure \ref{fig:simulation_test_result}. The agent's objective is to reach its target while avoiding obstacles within the environment. In Figure \ref{fig:simulation_test_result}b, we ensured that the distance between each static obstacle and the robot is greater than \(2 m\), allowing the robot to navigate between obstacles towards its goal without breaching the minimum collision distance of \(0.7 m\). We observed that all test models successfully navigated to their goals under this environment setup, with our model achieving the smoothest path and shortest travel time. Conversely, in the environment depicted in Figure \ref{fig:simulation_test_result2}, the TD3 baseline model failed as one of the obstacles was positioned approximately \(1.1 m\) from the robot's starting point, differing from its training experience, leading it to collide after a few steps. This issue did not occur with our model and TD3DT, as both models avoided entering the collision threshold by setting a \(1m\) threshold to trigger model retraining.   

During the retraining phase, TD3DT operates in a simulated environment starting from the robot's position at the initiation of retraining to the intended target. After successfully determining a path, it updates the pre-trained model and proceeds with navigation in the digital twin. While this approach is similar to our proposed model, we incorporate human expert intervention to control the robot's initial steps before handing over control to the autonomous system. 

Our experimental results, presented in Table \ref{tab:performance}, reveal significant insights into the performance of our proposed methodology compared with TD3 and TD3DT algorithms. The rate of success was markedly higher for our methodology, achieving a success rate of \(85\%\), compared to \(65\%\) and \(70\%\) for TD3 and TD3DT respectively. This suggests a robust capability of our approach in navigating complex environments.
\begin{table}[!t]
\caption{Summary of Performance Metrics\label{tab:performance}}
\centering
\begin{tabular}{|c|c|c|c|c|c|}
\hline
Metric & Proposed Method & TD3 & TD3DT & Improvement \\
\hline
Success Rate & 85\% & 65\% & 70\% & +15\%\\
\hline
Collision Rate & 10\% & 25\% & 20 & -10\%\\
\hline
Timeout Rate & 5\% & 10\% & 10\%& -5\%\\
\hline
Average Travel Time & 10sec & 17sec & 14sec& -6sec \\
\hline
\end{tabular}
\end{table}
Regarding safety metrics, our methodology demonstrated a lower collision rate at \(10\%\), compared to \(20\%\) for TD3 and \(25\%\) for TD3DT, underscoring enhanced navigational safety. Similarly, timeouts were less frequent in our trials, indicating more efficient task management under our proposed model.

Travel time also favoured our approach, with our methodology showing a 3-minute reduction compared to TD3DT during retraining. Also, with our model improving over time, it was able to achieve an average travel time of 6 seconds reduction per task compared to baseline models. This improvement not only illustrates faster task completion but also greater overall efficiency during navigation.
\subsection{Results in Real-World Testing}
In our research, we address a critical challenge in AI development: the gap between simulated testing and real-world application. This gap often presents significant hurdles due to the unpredictable nature of real-world environments that simulations may not fully capture. 

Our primary objective is to investigate and validate the performance of our proposed method in real-world conditions, thus enhancing the reliability of AI model transfer from controlled simulations to dynamic real-world settings. We conducted extensive testing, first in a simulated environment designed to mimic real-world variables as closely as possible, followed by direct deployment in real-world scenarios.

\begin{figure*}[!t]
\centering
\subfloat[]{\includegraphics[width=2.5in]{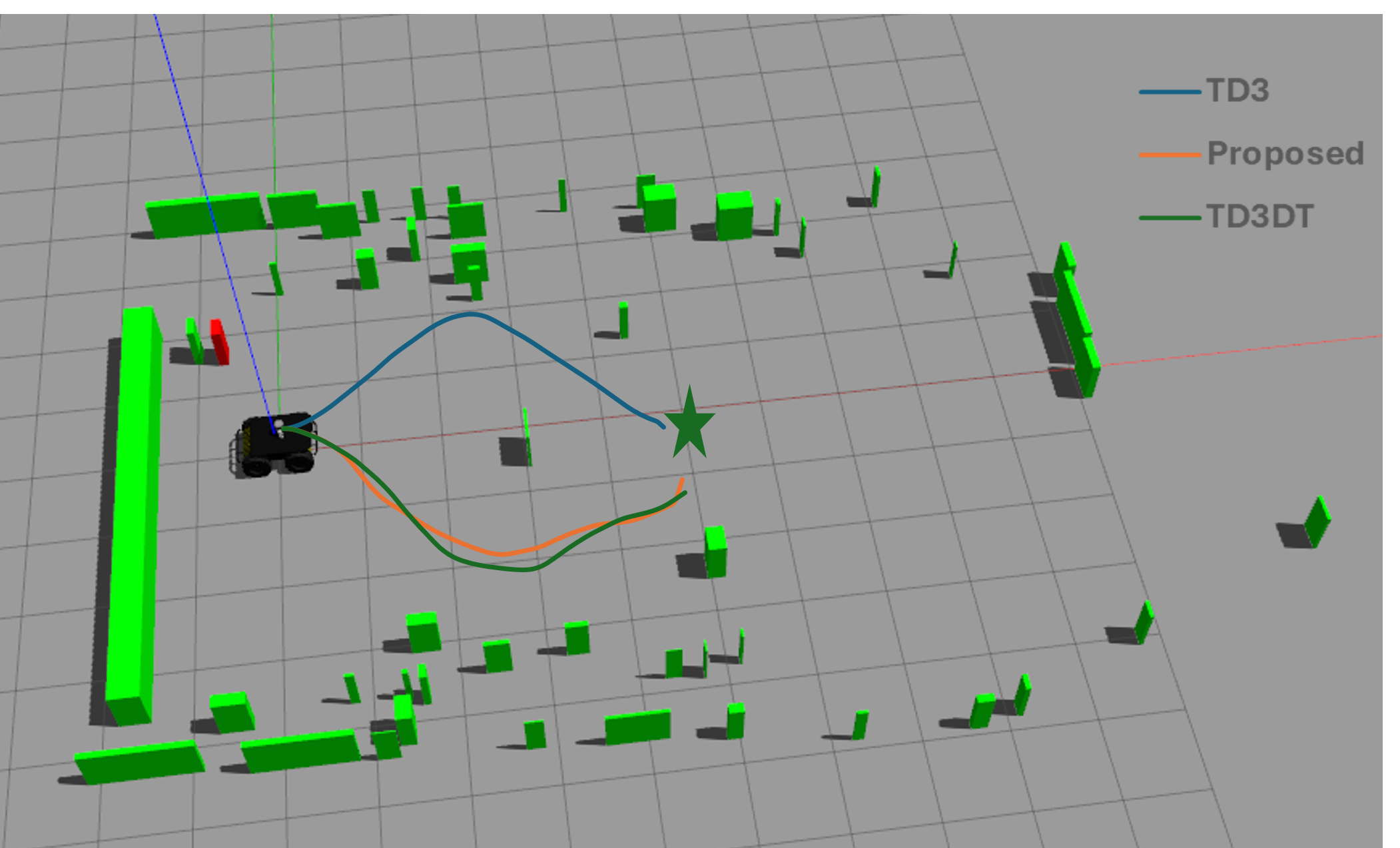}}%
\label{fig:realresult1}
\hfil
\subfloat[]{\includegraphics[width=2.5in]{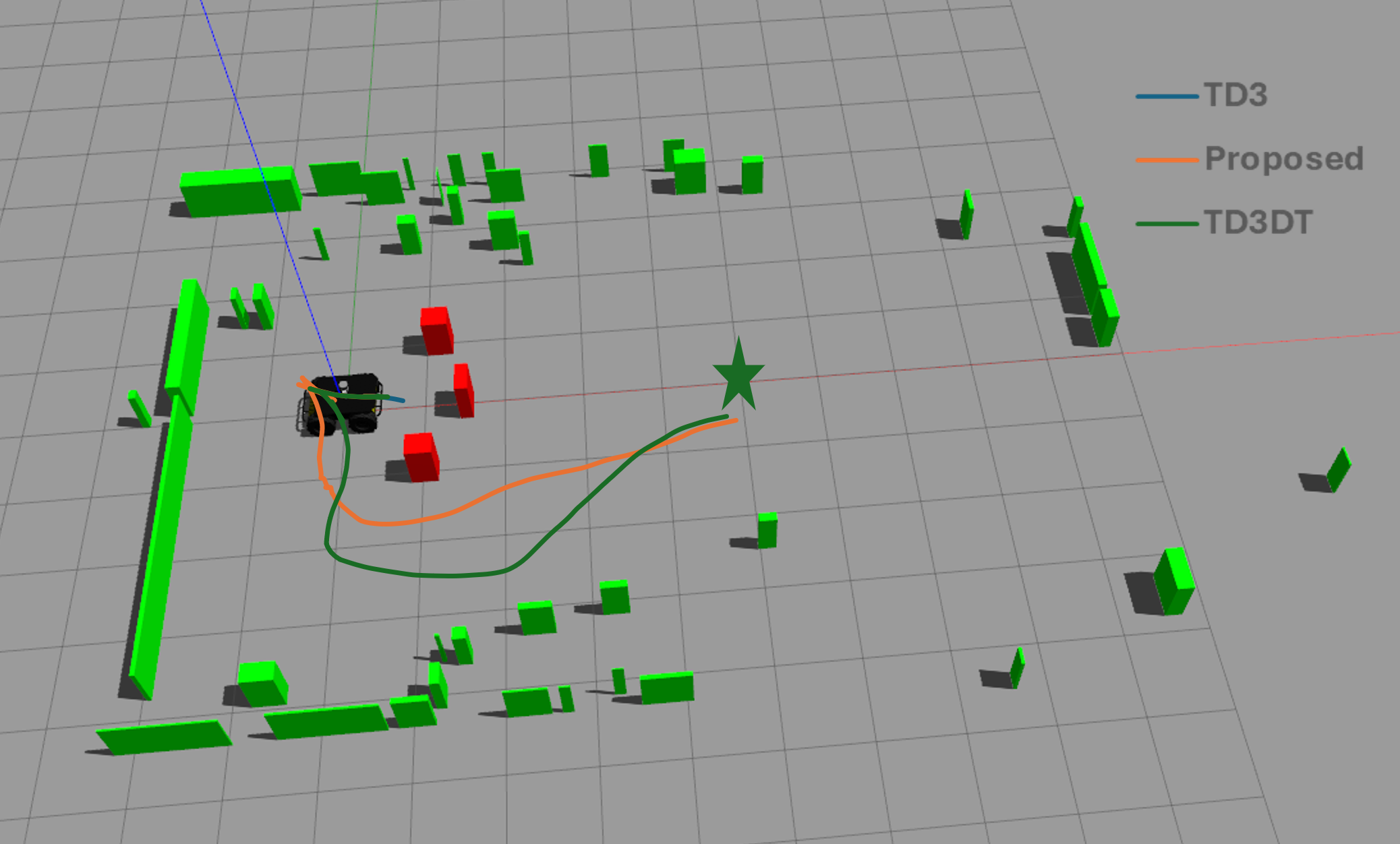}%
\label{fig:realresult2}}
\caption{Result of our proposed model against baseline models in the real world. (a) In this setup, all models had a positive success rate. (b) However, in this setup, both TD3DT and our proposed model reached their target with ours having a smoother path and reduced navigation time}.
\label{fig:realresult}
\end{figure*}

As shown in Figure \ref{fig:realresult}, the results from our real-world tests closely align with those observed in the simulation. This congruence is indicative of our method's robustness and its capability to handle real-world unpredictability effectively. Such outcomes not only validate our simulation design but also reinforce the efficacy of our model in practical applications.

\section{Conclusion}
\label{sec:conclusion}
In this paper, we present a novel digital-twin framework for RL real-time self-learning after deployment to further improve UGV's navigation problems. We enable the RL model to initially train autonomously in simulation to enable self-learning and reliance on human experts. During the pre-training, experiences are stored in a replay buffer and stores prioritized actions to a priority buffer. Also, in cases where the existing model underperforms after deployment, we enable human experts to provide real-time input to improve the model. Unlike the traditional method of deployment on the physical UGV, the deployment and control of the physical UGV is done in twin. We tested the performance of our model in both simulation and real-world ensuring our navigation systems are not only effective but also robust and safe for everyday use.

In developing a digital twin for a robotic system, support a robust framework where real-world operations and virtual simulations complement each other. This dual operation allows for safer, more efficient, and more rapid development and testing cycles, enhancing both the performance and reliability of the robotic system. The consistency of our model’s performance across both simulated and real-world tests not only validates our current methodologies but also sets the stage for future advancements in AI model training and deployment

\bibliographystyle{IEEEtran}
\bibliography{root}

\begin{thebibliography}{10}
\providecommand{\url}[1]{#1}
\csname url@samestyle\endcsname
\providecommand{\newblock}{\relax}
\providecommand{\bibinfo}[2]{#2}
\providecommand{\BIBentrySTDinterwordspacing}{\spaceskip=0pt\relax}
\providecommand{\BIBentryALTinterwordstretchfactor}{4}
\providecommand{\BIBentryALTinterwordspacing}{\spaceskip=\fontdimen2\font plus
\BIBentryALTinterwordstretchfactor\fontdimen3\font minus \fontdimen4\font\relax}
\providecommand{\BIBforeignlanguage}[2]{{%
\expandafter\ifx\csname l@#1\endcsname\relax
\typeout{** WARNING: IEEEtran.bst: No hyphenation pattern has been}%
\typeout{** loaded for the language `#1'. Using the pattern for}%
\typeout{** the default language instead.}%
\else
\language=\csname l@#1\endcsname
\fi
#2}}
\providecommand{\BIBdecl}{\relax}
\BIBdecl

\bibitem{hu2023use}
X.~Hu and R.~H. Assaad, ``The use of unmanned ground vehicles and unmanned aerial vehicles in the civil infrastructure sector: Applications, robotic platforms, sensors, and algorithms,'' \emph{Expert Systems with Applications}, p. 120897, 2023.

\bibitem{lv2021research}
J.~Lv, C.~Qu, S.~Du, X.~Zhao, P.~Yin, N.~Zhao, and S.~Qu, ``Research on obstacle avoidance algorithm for unmanned ground vehicle based on multi-sensor information fusion,'' \emph{Mathematical Biosciences and Engineering}, vol.~18, no.~2, pp. 1022--1039, 2021.

\bibitem{Nguyen}
M.-N. Nguyen, M.~Van, S.~McIlvanna, Y.~Sun, J.~Close, K.~Olayemi, and Y.~Jin, ``Model-free safety critical model predictive control for mobile robot in dynamic environments,'' \emph{IEEE Transactions on Intelligent Vehicles}, pp. 1--12, 2024.

\bibitem{rone2013mapping}
W.~Rone and P.~Ben-Tzvi, ``Mapping, localization and motion planning in mobile multi-robotic systems,'' \emph{Robotica}, vol.~31, no.~1, pp. 1--23, 2013.

\bibitem{fethi2018simultaneous}
D.~Fethi, A.~Nemra, K.~Louadj, and M.~Hamerlain, ``Simultaneous localization, mapping, and path planning for unmanned vehicle using optimal control,'' \emph{Advances in Mechanical Engineering}, vol.~10, no.~1, p. 1687814017736653, 2018.

\bibitem{wu2020robust}
Y.~Wu, Y.~Li, W.~Li, H.~Li, and R.~Lu, ``Robust lidar-based localization scheme for unmanned ground vehicle via multisensor fusion,'' \emph{IEEE transactions on neural networks and learning systems}, vol.~32, no.~12, pp. 5633--5643, 2020.

\bibitem{yang2016survey}
L.~Yang, J.~Qi, D.~Song, J.~Xiao, J.~Han, Y.~Xia \emph{et~al.}, ``Survey of robot 3d path planning algorithms,'' \emph{Journal of Control Science and Engineering}, vol. 2016, 2016.

\bibitem{shen2023targeted}
C.~Shen and G.~S. Soh, ``Targeted sampling dwa: A path-aware dwa sampling strategy for omni-directional robots,'' in \emph{International Design Engineering Technical Conferences and Computers and Information in Engineering Conference}, vol. 87363.\hskip 1em plus 0.5em minus 0.4em\relax American Society of Mechanical Engineers, 2023, p. V008T08A061.

\bibitem{cadena2016past}
C.~Cadena, L.~Carlone, H.~Carrillo, Y.~Latif, D.~Scaramuzza, J.~Neira, I.~Reid, and J.~J. Leonard, ``Past, present, and future of simultaneous localization and mapping: Toward the robust-perception age,'' \emph{IEEE Transactions on robotics}, vol.~32, no.~6, pp. 1309--1332, 2016.

\bibitem{macario2022comprehensive}
A.~Macario~Barros, M.~Michel, Y.~Moline, G.~Corre, and F.~Carrel, ``A comprehensive survey of visual slam algorithms,'' \emph{Robotics}, vol.~11, no.~1, p.~24, 2022.

\bibitem{aggarwal2020path}
S.~Aggarwal and N.~Kumar, ``Path planning techniques for unmanned aerial vehicles: A review, solutions, and challenges,'' \emph{Computer communications}, vol. 149, pp. 270--299, 2020.

\bibitem{sivashangaran2021intelligent}
S.~Sivashangaran and M.~Zheng, ``Intelligent autonomous navigation of car-like unmanned ground vehicle via deep reinforcement learning,'' \emph{IFAC-PapersOnLine}, vol.~54, no.~20, pp. 218--225, 2021.

\bibitem{9468918}
H.~Hu, K.~Zhang, A.~H. Tan, M.~Ruan, C.~Agia, and G.~Nejat, ``A sim-to-real pipeline for deep reinforcement learning for autonomous robot navigation in cluttered rough terrain,'' \emph{IEEE Robotics and Automation Letters}, vol.~6, no.~4, pp. 6569--6576, 2021.

\bibitem{henderson2018deep}
P.~Henderson, R.~Islam, P.~Bachman, J.~Pineau, D.~Precup, and D.~Meger, ``Deep reinforcement learning that matters,'' in \emph{Proceedings of the AAAI conference on artificial intelligence}, vol.~32, no.~1, 2018.

\bibitem{kirkpatrick2017overcoming}
J.~Kirkpatrick, R.~Pascanu, N.~Rabinowitz, J.~Veness, G.~Desjardins, A.~A. Rusu, K.~Milan, J.~Quan, T.~Ramalho, A.~Grabska-Barwinska \emph{et~al.}, ``Overcoming catastrophic forgetting in neural networks,'' \emph{Proceedings of the national academy of sciences}, vol. 114, no.~13, pp. 3521--3526, 2017.

\bibitem{mirowski2018learning}
P.~Mirowski, M.~Grimes, M.~Malinowski, K.~M. Hermann, K.~Anderson, D.~Teplyashin, K.~Simonyan, A.~Zisserman, R.~Hadsell \emph{et~al.}, ``Learning to navigate in cities without a map,'' \emph{Advances in neural information processing systems}, vol.~31, 2018.

\bibitem{olayemi2023impact}
K.~B. Olayemi, M.~Van, S.~McLoone, S.~McIlvanna, Y.~Sun, J.~Close, and N.~M. Nguyen, ``The impact of lidar configuration on goal-based navigation within a deep reinforcement learning framework,'' \emph{Sensors}, vol.~23, no.~24, p. 9732, 2023.

\bibitem{zhang2019reinforcement}
P.~Zhang, L.~Xiong, Z.~Yu, P.~Fang, S.~Yan, J.~Yao, and Y.~Zhou, ``Reinforcement learning-based end-to-end parking for automatic parking system,'' \emph{Sensors}, vol.~19, no.~18, p. 3996, 2019.

\bibitem{article}
T.~Hester, M.~Vecerik, O.~Pietquin, M.~Lanctot, T.~Schaul, B.~Piot, A.~Sendonaris, G.~Dulac-Arnold, I.~Osband, J.~Agapiou, J.~Leibo, and A.~Gruslys, ``Learning from demonstrations for real world reinforcement learning,'' 04 2017.

\bibitem{vecerik2018leveraging}
M.~Vecerik, T.~Hester, J.~Scholz, F.~Wang, O.~Pietquin, B.~Piot, N.~Heess, T.~Rothörl, T.~Lampe, and M.~Riedmiller, ``Leveraging demonstrations for deep reinforcement learning on robotics problems with sparse rewards,'' 2018.

\bibitem{Arakawa2018DQNTAMERHR}
\BIBentryALTinterwordspacing
R.~Arakawa, S.~Kobayashi, Y.~Unno, Y.~Tsuboi, and S.~ichi Maeda, ``Dqn-tamer: Human-in-the-loop reinforcement learning with intractable feedback,'' \emph{ArXiv}, vol. abs/1810.11748, 2018. [Online]. Available: \url{https://api.semanticscholar.org/CorpusID:53105957}
\BIBentrySTDinterwordspacing

\bibitem{wu2023toward}
J.~Wu, Z.~Huang, Z.~Hu, and C.~Lv, ``Toward human-in-the-loop ai: Enhancing deep reinforcement learning via real-time human guidance for autonomous driving,'' \emph{Engineering}, vol.~21, pp. 75--91, 2023.

\bibitem{wu2023human}
J.~Wu, Y.~Zhou, H.~Yang, Z.~Huang, and C.~Lv, ``Human-guided reinforcement learning with sim-to-real transfer for autonomous navigation,'' \emph{IEEE Transactions on Pattern Analysis and Machine Intelligence}, 2023.

\bibitem{luo2023human}
B.~Luo, Z.~Wu, F.~Zhou, and B.-C. Wang, ``Human-in-the-loop reinforcement learning in continuous-action space,'' \emph{IEEE Transactions on Neural Networks and Learning Systems}, 2023.

\bibitem{9645287}
R.~Cimurs, I.~H. Suh, and J.~H. Lee, ``Goal-driven autonomous exploration through deep reinforcement learning,'' \emph{IEEE Robotics and Automation Letters}, vol.~7, no.~2, pp. 730--737, 2022.

\bibitem{olayemi2024twin}
K.~Olayemi, M.~Van, S.~McLoone, Y.~Sun, J.~Close, N.~M. Nhat, and S.~McIlvanna, ``A twin delayed deep deterministic policy gradient algorithm for autonomous ground vehicle navigation via digital twin perception awareness,'' \emph{arXiv preprint arXiv:2403.15067}, 2024.

\bibitem{tao2022digital}
F.~Tao, B.~Xiao, Q.~Qi, J.~Cheng, and P.~Ji, ``Digital twin modeling,'' \emph{Journal of Manufacturing Systems}, vol.~64, pp. 372--389, 2022.

\bibitem{singh2021digital}
M.~Singh, E.~Fuenmayor, E.~P. Hinchy, Y.~Qiao, N.~Murray, and D.~Devine, ``Digital twin: Origin to future,'' \emph{Applied System Innovation}, vol.~4, no.~2, p.~36, 2021.

\bibitem{bottjer2023review}
T.~B{\"o}ttjer, D.~Tola, F.~Kakavandi, C.~R. Wewer, D.~Ramanujan, C.~Gomes, P.~G. Larsen, and A.~Iosifidis, ``A review of unit level digital twin applications in the manufacturing industry,'' \emph{CIRP Journal of Manufacturing Science and Technology}, vol.~45, pp. 162--189, 2023.

\bibitem{alhmiedat2023slam}
T.~Alhmiedat, A.~M. Marei, W.~Messoudi, S.~Albelwi, A.~Bushnag, Z.~Bassfar, F.~Alnajjar, and A.~O. Elfaki, ``A slam-based localization and navigation system for social robots: The pepper robot case,'' \emph{Machines}, vol.~11, no.~2, p. 158, 2023.

\bibitem{kastner2023predicting}
L.~K{\"a}stner, A.~Christian, R.~S. Mello, B.~Li, B.~Fatloun, and J.~Lambrecht, ``Predicting navigational performance of dynamic obstacle avoidance approaches using deep neural networks,'' in \emph{2023 32nd IEEE International Conference on Robot and Human Interactive Communication (RO-MAN)}.\hskip 1em plus 0.5em minus 0.4em\relax IEEE, 2023, pp. 1764--1771.

\bibitem{golchoubian2024uncertainty}
M.~Golchoubian, M.~Ghafurian, K.~Dautenhahn, and N.~L. Azad, ``Uncertainty-aware drl for autonomous vehicle crowd navigation in shared space,'' \emph{arXiv preprint arXiv:2405.13969}, 2024.

\end{thebibliography}
\end{document}